%Paper: astro-ph/9412095
%From: <moffat@medb.physics.utoronto.ca>
%Date: Wed, 28 Dec 1994 14:32:10 -0500

%Macro file for processing plain TeX
\magnification=1200
\voffset=0 true mm
\hoffset=0 true in
\hsize=6.5 true in
\vsize=8.5 true in
\normalbaselineskip=13pt
\def\doublespace{\baselineskip=20pt plus 3pt\message{double space}}
\def\singlespace{\baselineskip=13pt\message{single space}}
\let\spacing=\singlespace
\parindent=1.0 true cm

% bold face mathe italic fonts in dir 2160, 1800, 1643, and 1500
 %ambi in VAX

% also available in dir 1000,1095,1200,1315, and 1440

\newcount\equationumber \newcount\sectionumber
\sectionumber=1 \equationumber=1
\def\setsection{\global\advance\sectionumber by1 \equationumber=1}

\def\numbe{{{\number\sectionumber}{.}\number\equationumber}
                            \global\advance\equationumber by1}
\def\numberit{\eqno{(\number\equationumber)} \global\advance\equationumber by1}

\def\numberal{(\number\equationumber)\global\advance\equationumber by1}

\def\ccf#1{\,\vcenter{\normalbaselines
    \ialign{\hfil$##$\hfil&&$\>\hfil ##$\hfil\crcr
      \mathstrut\crcr\noalign{\kern-\baselineskip}
      #1\crcr\mathstrut\crcr\noalign{\kern-\baselineskip}}}\,}
\def\scf#1{\,\vcenter{\baselineskip=9pt
    \ialign{\hfil$##$\hfil&&$\>\hfil ##$\hfil\crcr
      \vphantom(\crcr\noalign{\kern-\baselineskip}
      #1\crcr\mathstrut\crcr\noalign{\kern-\baselineskip}}}\,}

\def\small3j#1#2#3#4#5#6{\def\st{\scriptstyle} % 3j-symbol - small size
   \bigl(\scf{\st#1&\st#2&\st#3\cr
           \st#4&\st#5&\st#6\cr} \bigr)}

   %Name of a nucleus

%\def\slashA{\hbox{$A\mkern-9mu/\mkern 9mu$}}

%\def\slasshA{\hbox{$A\mkern-9mu/\mkern 5mu$}}

\def\ref#1{$^{#1)}$}

   %Figure caption
              %#4 for caption

%...... subscripts and supscripts .....................................
\def\upa#1{\raise 1pt\hbox{\sevenrm #1}}
\def\dna#1{\lower 1pt\hbox{\sevenrm #1}}
\def\dnb#1{\lower 2pt\hbox{$\scriptstyle #1$}}
\def\dnc#1{\lower 3pt\hbox{$\scriptstyle #1$}}
\def\upb#1{\raise 2pt\hbox{$\scriptstyle #1$}}
\def\upc#1{\raise 3pt\hbox{$\scriptstyle #1$}}
\def\hprime{\raise 2pt\hbox{$\scriptstyle \prime$}}
\def\ccom{\,\raise2pt\hbox{,}}

%.... special maths symbols

\def\asymptotically#1{\;\rlap{\lower 4pt\hbox to 2.0 true cm{
    \hfil\sevenrm  #1 \hfil}}
   \hbox{$\relbar\joinrel\relbar\joinrel\relbar\joinrel
     \relbar\joinrel\relbar\joinrel\longrightarrow\;$}}
\def\Asymptotically#1{\;\rlap{\lower 4pt\hbox to 3.0 true cm{
    \hfil\sevenrm  #1 \hfil}}
   \hbox{$\relbar\joinrel\relbar\joinrel\relbar\joinrel\relbar\joinrel
     \relbar\joinrel\relbar\joinrel\relbar\joinrel\relbar\joinrel
     \relbar\joinrel\relbar\joinrel\longrightarrow$\;}}

\catcode`@=11
\def\C@ncel#1#2{\ooalign{$\hfil#1\mkern2mu/\hfil$\crcr$#1#2$}}
\def\gf#1{\mathrel{\mathpalette\c@ncel#1}}      % slash a small letter
\def\Gf#1{\mathrel{\mathpalette\C@ncel#1}}      % slash a big letter

\def\gapx{\lower 2pt \hbox{$\buildrel>\over{\scriptstyle{\sim}}$}}
\def\lapx{\lower 2pt \hbox{$\buildrel<\over{\scriptstyle{\sim}}$}}

\def\nablaleft{\hbox{\raise 6pt\rlap{{\kern-1pt$\leftarrow$}}{$\nabla$}}}
\def\nablaright{\hbox{\raise 6pt\rlap{{\kern-1pt$\rightarrow$}}{$\nabla$}}}
\def\nablaboth{\hbox{\raise 6pt\rlap{{\kern-1pt$\leftrightarrow$}}{$\nabla$}}}

\def\boks#1#2{{\hsize=#1 true cm\parindent=0pt
  {\vbox{\hrule height1pt \hbox
    {\vrule width1pt \kern3pt\raise 3pt\vbox{\kern3pt{#2}}\kern3pt
    \vrule width1pt}\hrule height1pt}}}}

\def\heading{ }
\def\range{ }

\def\body{\vfill\eject\parindent=1.0 true cm
 \ifx\spacing\singlespace\singlespace\else\doublespace\fi}
\def\title#1{\centerline{{\bf #1}}}

\def\today{\ifcase\month\or
  January\or February\or March\or April\or May\or June\or
  July\or August\or September\or October\or November\or December\fi
  \space\number\day, \number\year}
\let\date=\today
\newcount\hour \newcount\minute
\countdef\hour=56
\countdef\minute=57
\hour=\time
  \divide\hour by 60
  \minute=\time
  \count58=\hour
  \multiply\count58 by 60
  \advance\minute by -\count58

\def\sectionskip{\penalty-500\vskip24pt plus12pt minus6pt}

\def\sec{\hbox{\lower 1pt\rlap{{\sixrm S}}{\hbox{\raise 1pt\hbox{\sixrm S}}}}}
\def\section#1\par{\goodbreak\message{#1}
    \sectionskip\nobreak\noindent{\bf #1}\vskip0.3cm \noindent}

\nopagenumbers
\headline={\ifnum\pageno=\count31\frontheadline
  \else{\ifnum\pageno=0\frontheadline
     \else{{\raise 2pt\hbox to \hsize{\paperhead}}}\fi}\fi}
%\headline={\ifnum\pageno=\count31\frontheadline
%  \else{\ifnum\pageno=0\frontheadline
%     \else{\underbar{\raise 2pt\hbox to \hsize{\paperhead}}}\fi}\fi}

\footline={\centerline{\sevenbf \folio}}
\def\frontheadline{\sevenbf \hfil}
\def\paperhead{\sevenbf \heading\ \range\hfil\folio}
\newdimen\pagewidth \newdimen\pageheight \newdimen\ruleht
\maxdepth=2.2pt
\pagewidth=\hsize \pageheight=\vsize \ruleht=.5pt

\def\onepageout#1{\shipout\vbox{ % here we define one page of output
    \offinterlineskip % butt the boxes together
  \makeheadline
    \vbox to \pageheight{
         #1 % now insert the main information
 \ifnum\pageno=\count31{\vskip 21pt\line{\the\footline}}\fi
 \ifvoid\footins\else %footnot ino is present
 \vskip\skip\footins \kern-3pt
 \hrule height\ruleht width\pagewidth \kern-\ruleht \kern3pt
 \unvbox\footins\fi
 \boxmaxdepth=\maxdepth}
 \advancepageno}}
\output{\onepageout{\pagecontents}}
\count31=-1
\topskip 0.7 true cm
%end of TeX macro file
\doublespace
\centerline{\bf Galaxy Dynamics Predictions in the Nonsymmetric Gravitational
Theory}
\centerline{\bf J. W. Moffat}
\centerline{\bf Department of Physics}
\centerline{\bf University of Toronto}
\centerline{\bf Toronto, Ontario M5S 1A7, Canada}
\vskip 0.2 true in
\centerline{\bf Abstract}
\vskip 0.2 true in
In the weak field approximation, the nonsymmetric gravitational theory has,
in addition to the Newtonian gravitational potential,
a Yukawa potential produced by the exchange of
a spin $1^+$ boson between fermions. If the range $r_0$ is of order $30$ kpc,
then the potential due to the interaction of known neutrinos in the halos
of galaxies can explain the flat rotation curves of galaxies. The results
are based on a physical linear approximation to the NGT field equations
and they are consistent with equivalence principle observations, other
solar system gravitational experiments and the binary pulsar data.
\vskip 2 true in
{\bf UTPT 94-39}
\vskip 0.2 true in
e-mail address: moffat@medb.physics.utoronto.ca
\par\vfil\eject
After two decades there has not been any observation of
exotic dark matter candidates. Recent observational results using the HST
have excluded faint stars as a source of dark matter in the solar
neighborhood$^{1}$. However, the galaxy dynamics observations continue to
pose a serious challenge to gravitational theories. The data are in sharp
contradiction with Newtonian dynamics, for virtually all spiral galaxies
have rotational velocity curves which tend towards a constant value$^{2}$.
Similar results are observed in gravitational lensing$^{3}$.

As in the case of anomaly problems in solar dynamics of the past century,
concerning Uranus and Mercury, there are two ways to circumvent the
problem. The most popular is to postulate the existence of dark matter$^{4}$.
It is assumed that dark matter exists in massive almost spherical halos
surrounding galaxies. About 90\% of the mass is in the form of dark matter
and this can explain the flat rotational velocity curves of galaxies.
However, the scheme is not economical, because it requires two or three
parameters to describe different kinds of galactic systems and no satisfactory
model of galactic halos is known.

The other possible explanation for the galactic observations is to say that
Newtonian gravity is not valid at galactic scales. This has been the
subject of much discussion in recent years$^{5-9}$. We know that Einstein's
gravitational theory (EGT) correctly describes solar system observations
and the observations of the binary pulsar PSR 1913+16$^{10}$.
Therefore, any explanation of galactic dynamics based on gravity must be
contained in a modified gravitational theory that is consistent with EGT.
The following constraints on a classical gravitational theory are:

{\smallskip\obeylines
(1) The theory must be generally covariant i.e., the field equations should
be independent of general coordinate transformations and should reduce to
special relativity dynamics in flat Minkowskian spacetime.

(2) The theory should be derivable from a least action principle in order to
guarantee the consistency of the theory.

(3) The linear approximation should be consistent i.e., there should not be
any ghost poles, tachyons or higher-order poles and the asymptotic flat space
boundary conditions should be satisfied.

(4) The equations of motion of test particles should be consistent with local
equivalence principle tests.

(5) All solar system tests of gravity and the observed rate of decay of the
binary pulsar should be predicted by the theory.\smallskip}

We shall now consider the predictions of a new version of the nonsymmetric
gravitational theory which can satisfy all the above criteria$^{11-15}$.
The theory has a linear approximation free of ghost poles, tachyons and
higher-order poles with field equations for a massive spin $1^+$ boson with
a range parameter, $\mu^{-1}=r_0$, corresponding to Proca-type equations for an
antisymmetric potential. The expansion of the field equations about an
arbitrary EGT background metric is also consistent and satisfies the
physical boundary conditions at asymptotically flat infinity.

An important result of the theory is that the field equations have a
spherically
symmetric static solution, which is completely regular everywhere in spacetime
and possesses no black hole event horizons$^{16,17}$. Black holes are replaced
by superdense objects (SDO's) which do not have null surfaces.

A derivation of the equations of motion of test particles yields
the following potential in the weak field approximation$^{14}$:
$$
V(r)=-{G_{\infty}M\over r}\biggl(1-{g^2\over G_{\infty}m_tM}Y_tY_s
e^{-\mu r}\biggr),
\numberit
$$
where $g^2$ is a coupling constant measuring the strength of the coupling
of the antisymmetric field $g_{[\mu\nu]}$ to matter and $Y_t$ and $Y_s$
denote the NGT charges of the test particle and the source, respectively.
Moreover, $G_{\infty}$ is the gravitational constant at large distances.
For $r << r_0$, the Newtonian laws apply with the local gravitational
constant:
$$
G_0=G_{\infty}(1-{g^2\over G_{\infty}m_tM}Y_tY_s).
\numberit
$$
Since the exchanged boson is a spin $1^+$ particle, the Yukawa potential
term corresponds to a repulsive force.

We shall assume that the NGT charge is associated with fermion particles:
$$
Y=Y_B+Y_{\nu},
\numberit
$$
where $Y_B$ and $Y_{\nu}$ denote the baryon charge and neutrino charge,
respectively. Let us assume that the mass $M$ is dominated by the rest
mass of the constituents:
$$
M\approx m_N(N+Z)+m_{\nu}N_{\nu},
\numberit
$$
and that, in addition, we have
$$
M\approx m_N(N+Z),
\numberit
$$
where $m_N$, $m_{\nu}$ and $N_{\nu}$ denote the nucleon mass, the neutrino
mass and
the number of neutrinos, respectively. Also, $N$ and $Z$ denote the number
of neutrons and protons, respectively. Then, Eq.(1) can be
written as
$$
V(r)=-{G_{\infty}M\over r}\biggl[1-\eta_t Y_t\biggl({Y_s\over N+Z}\biggr)
e^{-\mu r}\biggr],
\numberit
$$
where $\eta_t=g^2/G_{\infty}m_tm_N$.

The rotational velocity is determined by
$$
v_c=\biggl({G_{\infty}M\over r}\biggr)^{1/2}
\biggl[1-(1+{r\over r_0})\eta_t Y_t\biggl({Y_s\over N+Z}\biggr)
e^{-r/r_0}\biggr]^{1/2}.
\numberit
$$
In terms of the neutron excess, we can define
$$
\rho={N-Z\over N+Z},
\numberit
$$
and we have$^{18}$
$$
{Y_B\over N+Z}={1\over \sqrt{2}}\biggl[\hbox{cos}(\theta
-{\pi\over 4})-\rho\hbox{sin}(\theta-{\pi\over 4})\biggr].
\numberit
$$

The best observational limits on violations of the equivalence principle
come from E\"otv\"os-type experiments$^{19,20}$ that measure the differential
acceleration of two bodies towards the Earth. These limit
$(\delta a/g)_{\oplus}$, where $\delta a=a_1-a_2$ ($a_i$ is the
measured acceleration of body $i$) and $g$ is the gravitational
acceleration at the Earth's surface. We have for $r_{\oplus} << r_0$:
$$
\biggl({\delta a\over g}\biggr)_{\oplus}\approx \eta_B\delta Y_i
\biggl({Y\over N+Z}\biggr)_{\oplus} < 10^{-12},
\numberit
$$
where $\eta_B=g^2/G_{\infty}m_N^2$. For the two examples, $\theta=\pm \pi/4$
for which $Y_{B\pm}=(N\pm Z)/\sqrt{2}$, we have that
$Y/(N+Z)\approx 10^{-2}$ for the Earth and $\delta Y_i$ ranges from 0 to
$10^{-1}$
for materials compared in the experiments for $Y_{B-}$, while for $Y_{B+}$
we have that ${Y/(N+Z)}\approx 1$ for the Earth and $\delta Y_i
\sim 10^{-3}$ for differences
in materials. Therefore, we conclude that $\eta_B < 10^{-9}$ in order not
to violate these accurate experiments. Other constraints coming from
perihelion-shift measurements and those from the binary pulsar
and classical binary star systems such as DI Herculis would also be
consistent with the latter bound on $\eta_B$$^{21}$.

Sanders$^{22,23}$ has done an extensive phenomenological analysis of fits
to the rotational velocity curves of galaxies, using a repulsive Yukawa
potential added to the attractive Newtonian potential:
$$
V(r)=-{G_{\infty}M\over r}\biggl(1-\alpha e^{-r/r_0}\biggr),
\numberit
$$
where $\mu^{-1}=r_0\approx 30$ kpc and $\alpha=0.92$. However, we see that
if we assume only a coupling, in NGT, to baryons, then such fits would
overwhelmingly violate the equivalence principle measurements, as noted
by Sanders$^{6}$. We shall instead assume that the dominant coupling
is due to neutrino pairs with the potential:
$$
V_G(r)=-{G_{\infty}M\over r}\biggl(1-\gamma_{\nu}e^{-r/r_0}\biggr),
\numberit
$$
where
$$
\gamma_{\nu}={g^2N_{\nu}\over G_{\infty}m_{\nu}m_N}\biggl({Y\over
N+Z}\biggr)_G.
\numberit
$$
The equivalence principle tests and other observational gravitational tests
no longer limit $\alpha=\gamma_{\nu}$ to small values. We can have
$\gamma_{\nu} \sim 0.92$
or larger values of $\gamma_{\nu}$, depending on the values of the constants
$m_{\nu}$ and $N_{\nu}$ and assuming a universal value for the NGT
fermion coupling constant $g^2$.

The present upper bound on $m_{\nu}$ for the
electron-dominated family is 9 ev and the lower bound still
includes zero$^{24}$. The present mean number density of
neutrinos plus their partners in one family, determined by the density of
relict neutrinos, is fixed by the cosmic background temperature to be$^{25}$:
$$
n_{\nu}=113\,\,\hbox{neutrinos}\,{\hbox{cm}}^{-3}.
\numberit
$$
A value for the neutrino mass can be obtained from estimates of the density
of neutrinos in galaxy halos$^{25}$:
$$
m_{\nu}={70\over [r_{1/2}(\hbox{kpc})]^{1/2}}
\biggl({200\,\hbox{km}\,s^{-1}\over v_c}\biggr)^{1/4},
\numberit
$$
where $r_{1/2}$ denotes the galaxy core radius.
For typical dark halos of giant galaxies, with $v_c\sim 200\,\hbox{km}\,s^{-1}$
and core radii $r_{1/2}$ of a few kiloparsecs, the neutrino mass obtained
from (15) is similar to the mass at which neutrinos close the universe
at an acceptable value of the Hubble constant. However, there are serious
problems with this scenario for dwarf spheroidal galaxies in the halo of
the Milky Way$^{25,26}$. For Draco and Ursa Minor, the resulting neutrino mass
is
$$
m_{\nu}\sim 400\,ev,
\numberit
$$
which is an order of magnitude or more above what is allowed by the mean
mass density (14).

Our fits to the galaxy rotational velocity curves are not restricted by
the problems with dwarf galaxies and we can obtain fits for $m_{\nu} \leq 1$ ev
for reasonable values of $N_{\nu}$. We note that
$m_{\nu} \geq 10^9 m_N$ which means that we gain a factor of $10^9$ or
more in Eq.(13), when compared to the baryon coupling contribution.

Since $N_{\nu}$ can vary from galaxy to galaxy, we should be able to fit
the observed fact that for low luminosity-low rotation velocity galaxies
the rotation curve still tends to be rising at the last measured points,
whereas in high rotation velocity galaxies the opposite seems to be true
$^{6}$, i.e., the rotation curves are decreasing but still more slowly
than is expected from the light distributions. In the empirical work of
Sanders, based on Eq.(11), it was predicted that larger galaxies should exhibit
larger mass discrepancies, which does not seem necessarily to follow from the
data, e.g., for the very large galaxy UGC 2885, there is no evidence
of any mass discrepancy out to a radius of 60 or 70 kpc. In contrast,
there are very small galaxies such as the spiral UGC 2258, which
display a significant mass discrepancy at a radius less than 10 kpc.
Again this kind of behavior depends upon the composition of the neutrino
halo of the galaxy, and a fit to the data should be possible, although
further data fitting is necessary to confirm this scenario.

Regarding the dark matter problem at
cosmological scales a cosmological constant $\Lambda$ is one way
to make a low-density universe consistent with the condition from inflation
that space curvature is negligibly small, without invoking the hypothesis
of exotic dark matter. A positive $\Lambda$ tends to pull clusters apart,
but the effect can be ignored for interesting values of $\Lambda$. Perhaps,
the increased value of $G_{\infty}$, obtained in the present scheme
for very large distances, could improve the clustering effect
of large scale gas clouds and help to account for the formation of
galaxies.

{}From the predicted weak field gravitational potential of NGT and the work
of Sanders, we have
seen that it is possible to fit a wide class of galaxy rotational
velocity curves for fixed values of $g^2$ and  $\mu$ and from
variable light distributions and galaxy halo neutrino density
distributions. Values of $m_{\nu}$ are allowed that do not contradict
the experimental bounds on this constant and that are not inconsistent
with cosmological estimates of the mean density of neutrinos. A positive
feature of this scheme is that only the {\it known} neutrino dark matter
needs to be postulated to fit the data.
Moreover, the theory is generally covariant, posseses a physically
consistent linear approximation and allows a relativistic
calculation of the bending of light which agrees with the solar experiments.
\vskip 0.2 true in
{\bf Acknowledgement}
\vskip 0.2 true in
This work was supported by the Natural Sciences and Engineering Research
Council of Canada.
\vskip 0.2 true in
\centerline{\bf References}
\vskip 0.2 true in
\item{1.}{NASA News Release No.:STScI-PRC94-41a Hubble Space Telescope News,
November 15, 1994.}
\item{2.}{R. H. Sanders, Astron. Astrophys. Rev. {\bf 2}, 1 (1990).}
\item{3.}{A. Dar, Nucl. Phys. B (Proc. Suppl.) {\bf 28A}, 321 (1992).}
\item{4.}{K. M. Ashman, Publications of the Astron. Soc. of the Pacific
{\bf 104}, 1109 (1992).}
\item{5.}{J. D. Bekenstein, in Proc. 2nd Canadian Conf. on General
Relativity and Relativistic Astrophysics, ed. C. Dyer and T. Tupper
(Singapore, World Scientific, 1988), p.68.
\item{6.}{R. H. Sanders, Mon. Not. R. Aston. Soc. {\bf 223}, 539 (1986).}
\item{7.}{D. H. Eckhardt, Phys. Rev. D {\bf 48}, 3762 (1993).}
\item{8.}{P. D. Mannheim, Found. Phys. {\bf 24}, 487 (1994); P. D.
Mannheim and D. Kazanas, Gen. Relativ. Gravit. {\bf 26}, 337 (1994).}
\item{9.}{V. V. Zhytnikov and J. M. Nester, National Central University,
Chung-Li, Taiwan, preprint, gr-qc/9410002, 1994.}
\item{10.}{C. M. Will, Theory and Experiment in Gravitational Physics,
Cambridge University Press, Cambridge U.K. new edition 1993.}
\item{11.}{J. W. Moffat, Toronto preprint, UTPT-94-28,
gr-qc/9411006, 1994.}
\item{12.}{J. W. Moffat, Toronto preprint, UTPT-94-30,
gr-qc/9411027, 1994.}
\item{13.}{J. L\'egar\'e and J. W. Moffat, Toronto preprint,
UTPT-94-36, gr-qc/9412009, 1994.}
\item{14.}{J. L\'egar\'e and J. W. Moffat, Toronto preprint,
UTPT-94-38, gr-qc/9412074, 1994.}
\item{15.}{N. J. Cornish, in preparation.}
\item{16.}{N. J. Cornish and J. W. Moffat, Phys. Letts B {\bf 336},
337 (1994).}
\item{17.}{N. J. Cornish and J. W. Moffat, in press J. Math. Phys.}
\item{18.}{A. de R\'ujula, Phys. Lett. B {\bf 180}, 213 (1986).}
\item{19.}{C. W. Stubbs et al., Phys. Rev. Lett. {\bf 58}, 1070 (1987);
E. G. Adelberger et al., {\it ibid} {\bf 59}, 849 (1987).
For torsion balance experiments with $r_0 >> r_{\oplus}$, the experiment
can only detect the composition dependent force perpendicular to the
torsion wire, so there is an additional suppression factor $\sim 10^{-3}$.}
\item{20.}{T. M. Niebauer, M. P. McHugh, and J. E. Faller, Phys. Rev. Lett.
{\bf 59}, 609 (1987); in Neutrinos and Exotic Phenomena, Moriond Meetings
1988 and 1990. Editions Fronti\'eres, Gif-sur-Yvette, France.}
\item{21.}{C. P. Burgess and J. Cloutier, Phys. Rev. D {\bf 38}, 2944 (1988).}
\item{22.}{R. H. Sanders, Astron. Astrophys. {\bf 136}, L21-L23 (1984).}
\item{23.}{R. H. Sanders, Astron. Astrophys. {\bf 154}, 135-144 (1986).}
\item{24.}{Review of Particle Properties, Phys. Rev. D {\bf 50}, 1173
(1994).}
\item{25}{For references and a review, see: P. J. E. Peebles, Principles of
Physical Cosmology, Princeton University Press, Princeton, N. J. 1993.}
\item{26.}{O. E. Gerhard and D. N. Spergel, Ap. J. {\bf 389}, L9 (1992).}

\end